\documentclass[sigconf,screen]{acmart}

\usepackage{algorithmic}
\usepackage{graphicx}
\usepackage{textcomp}
\usepackage{tcolorbox}
\usepackage{pifont}
\usepackage{xcolor}
\usepackage{soul}
\usepackage{booktabs}
\usepackage{multirow}
\usepackage{float}
\usepackage{url}
\usepackage{graphicx}
\usepackage{balance}
\usepackage{microtype}
\usepackage[numbered]{bookmark}
\usepackage{tcolorbox}
\usepackage{titlesec}
\usepackage{enumitem}
\usepackage{seqsplit}

\setlist[itemize,enumerate]{noitemsep, topsep=0pt, leftmargin=1.0em}
\hypersetup{
	pdfborder          = {0 0 0},
	breaklinks         = true,
	bookmarksnumbered  = true,
	pdfsubject         = {1st International Workshop on Natural Language-based Software Engineering (NLBSE 2022)},
	pdfkeywords        = {Program Comprehension} {Identifier Names} {Rename Refactoring} ,
	pdftitle           = {Understanding Digits in Identifier Names: An Exploratory Study},
	pdfauthor          = {Anthony Peruma and Christian D. Newman}
}

\acmConference[NLBSE '22]{Proceedings of the 1st International Workshop on Natural Language-based Software Engineering}{May 8, 2022}{Virtual Event, USA}

\renewcommand\footnotetextcopyrightpermission[1]{}
\usepackage{draftwatermark}
\SetWatermarkText{PREPRINT}
\SetWatermarkLightness{0.88}
\SetWatermarkScale{0.33}

\usepackage{listings}

\definecolor{javared}{rgb}{0.6,0,0} 
\definecolor{javagreen}{rgb}{0.25,0.5,0.35} 
\definecolor{javapurple}{rgb}{0.5,0,0.35} 
\definecolor{javadocblue}{rgb}{0.25,0.35,0.75} 
 
\newcommand{\RQA}{\textbf{RQ1}: How does identifier renaming operations in the source code impact the existence of digits in an identifier's name?}

\newcommand{\RQC}{\textbf{RQ2}: How do developers utilize digits in an identifier's name to convey meaning?}

\begin{document}



\title{Understanding Digits in Identifier Names: An Exploratory Study}

\author{Anthony Peruma}
\email{axp6201@rit.edu}
\affiliation{%
  \institution{Rochester Institute of Technology}
  \city{Rochester}
  \state{New York}
  \country{USA}
}

\author{Christian D. Newman}
\email{cnewman@se.rit.edu}
\affiliation{%
  \institution{Rochester Institute of Technology}
  \city{Rochester}
  \state{New York}
  \country{USA}
}

\begin{abstract}
Before any software maintenance can occur, developers must read the identifier names found in the code to be maintained. Thus, high-quality identifier names are essential for productive program comprehension and maintenance activities. With developers free to construct identifier names to their liking, it can be difficult to automatically reason about the quality and semantics behind an identifier name. Studying the structure of identifier names can help alleviate this problem. Existing research focuses on studying words within identifiers, but there are other symbols that appear in identifier names-- such as digits. This paper explores the presence and purpose of digits in identifier names through an empirical study of 800 open-source Java systems. We study how digits contribute to the semantics of identifier names and how identifier names that contain digits evolve over time through renaming. We envision our findings improving the efficiency of name appraisal and recommendation tools and techniques.
\end{abstract}

\maketitle

\section{Introduction}
\label{Section:introduction}
As lexical tokens that uniquely identify entities in the code (such as classes, methods, variables, etc.), identifier names play an essential part in program comprehension activities, with well-constructed names improving comprehension activities by an estimated 19\% \cite{Hofmeister2017SANER}. Ideally, for identifier names to assist developers with understanding the code, the name must be unambiguous, and intent-revealing in communicating the purpose and behavior of its associated source code \cite{Butler2009FSE}. However, with developers having the ability to craft names using a variety of terms \cite{Peruma2020ICPC,Newman2020JSS, droricse:2021}, it is challenging to ensure consistent quality of the identifiers in the source code. This is exacerbated by the fact that around 70\% \cite{Deissenboeck2006SQJ} of the characters in the code base are dedicated to identifier names. Ignorance of the quality of identifier names is ignorance of the quality of a large portion of the code. 

To correct low quality identifiers, developers rename \cite{Fowler:1999} them. Rename refactorings are defined as refactorings that modify the name of an identifier without modifying the intended behavior of the code. Many Integrated Developer Environments (IDEs) offer a built-in rename refactoring functionality. Most of these IDEs only support the mechanical act of renaming; they allow a developer to choose what identifier they want to rename, accept the new name should be used, and then perform checks to avoid name collisions. There is little or no support beyond this to assist developers in determining when and how (e.g., help them pick terms). This is in spite of the fact that renaming is one of the most common refactorings \cite{Peruma2018IWoR, Peruma2019MobileSoftAndroid}; further highlighting that naming is a severe problem in software maintenance.

To understand the characteristics associated with high-quality names, past work in this area has focused on both empirical and developer-centric studies. To this extent, research studies have surveyed professional developers \cite{Alsuhaibani2021ICSE}, examined the presence, significance, and influence of abbreviations \cite{Newman2019ICSMEa,Newman2019ICSMEb, Jiang:2021, Jiang:2022}, acronyms \cite{Butler2015ICSME}, the comprehensibility \cite{Hofmeister2017SANER,Beniamini2017ICPC,Schankin2018ICPC} of different types of names, and investigated the semantic evolution of the identifiers \cite{Arnaoudova2014TSE,Peruma2020JSSRenames}. However, these existing studies focused on the words that make up identifiers, not digits. Therefore, this study expands our understanding of identifier names by studying the digits that appear within them. Our work aims to help lay the foundation in this area of research by conducting an exploratory study on the presence and purpose of digits in an identifier's name.

\subsection{Goal \& Research Questions}
The goal of this study is to understand the structure of identifier names containing digits and the semantics expressed by digits in identifier names. To this extent, we study \textit{the part played by digits in identifier names} by examining the name's evolution and the meaning conveyed by the digit to the overall purpose of the identifier. We envision findings from our study supporting the development of tools and techniques in identifier name recommendation and appraisal and the auto-generation of comprehension-friendly source code. We answer the following research questions (RQs):

\vspace{3mm}
\textbf{\RQA}
This question explores the extent to which digits are present in identifier names and how they change over time. Findings from this RQ inform us of the volume and characteristics of such identifiers; including if digits are generally preserved after a rename has been applied, and the number of digits that typically occur in a name.

\vspace{3mm}
\textbf{\RQC}
In this RQ, we perform a qualitative examination of the terms in an identifier's name and its related code to determine the rationale behind the presence of digits in an identifier's name. Through our analysis, we establish a taxonomy for the presence of digits in an identifier's name.

\subsection{Contribution}
The main contributions from this work are as follows:
\begin{itemize}
    \item The results show that digits are frequently preserved when a name changes (i.e., renames), indicating that these digits may have a significant role in helping developers understand the meaning/purpose of the identifier.
    \item Digits found in identifier names play varying roles that we can detect by looking at how they are used and structured. We taxonomize these roles; providing definitions and examples of the different ways digits were used in identifier names.
    \item Our findings provide a strong, if initial, step towards further understanding the use of digits in identifier names. Our discussion provides a path for future work to further contribute to our understanding of identifier naming in general, and the use of special characters/words in identifier works in particular.
    \item A discussion of the research challenges that this research poses. Thereby helping to direct other research in productive directions that may further illuminate much-needed answers for how digits can help serve program comprehension and identifier name quality.
\end{itemize}


\section{Related Work}
\label{Section:related_work}
This section reports on work related to our study. We divide the works into two specific areas -- studies on identifier naming in terms of their quality and structure and studies that examine the quality of the names during their evolution.

\subsection{Identifier Naming}

\subsubsection*{Naming Practices:} 
A study on Java naming practices by Gresta et al. \cite{Gresta21Naming} identifies eight categories, including the presence of digits in the name. The authors highlight common identifier names and indicate that context usually plays a role in determining the name developers select for an identifier, with single-letter names commonly used in short-scope contexts. Work on identifier naming by Butler et al. include INVocD, a database of identifier name declarations and source code structure of 60 Java projects \cite{Butler2013MSR}. The authors use this dataset to construct and test a naming convention checking library for Java systems and conducted a study to examine the adherence to naming conventions in Java programs \cite{Butler2015ICSME}. An investigation into the occurrence of single-letter variable names by Beniamini et al. \cite{Beniamini2017ICPC} shows that such names account for 10-20\% of all names. The authors also show that these names do not significantly impact comprehension and that language specifics and domain influence the letter chosen by the developer. Newman et al. conducted an empirical study of abbreviations and expansions \cite{Newman2019ICSMEb}, producing a dataset of manually verified 850 abbreviation-expansion pairs \cite{Newman2019ICSMEa}. Work by Hofmeister et al.  \cite{Hofmeister2017SANER} on the comprehension of names containing letters and/or abbreviations shows that developer comprehension speed increased by 19\% when identifiers had only words instead of abbreviations or letters. Alsuhaibani et al. \cite{Alsuhaibani2021ICSE} survey developers for their opinions on method naming standards. The authors report that developers agree with the method naming standards proposed by the authors. A study by Lawrie et al. \cite{Lawrie2006ICPC} on single-letter, abbreviations, and full words identifier show that full word identifiers lead to the best comprehension. However, there were cases where there was no statistical difference between full words and abbreviations.

\subsubsection*{Grammar Patterns:} An empirical study by Newman et al. \cite{Newman2020JSS} studied identifier name patterns and abstracted them into high-level descriptions of their purpose and semantics, called grammar patterns. They created a catalogue of identifier naming patterns \cite{newman2021Catalogue}, based on this work, that show how part-of-speech templates, called Grammar Patterns, can be used to analyze the meaning of different types of identifier names. In examining test method names, Peruma et al. \cite{Peruma2020ICPC} highlight multiple grammar patterns developers utilize in naming test methods and show that these patterns do not change very often during rename activities. Wu and Clause \cite{Wu2020JSS} utilize grammar patterns to identify non-descriptive test method names and propose a set of test method name patterns. Zhang et al. \cite{Zhang2020Access} study general characteristics of a large corpus of identifier names. Utilizing the Stanford Parser, one of their results shows that nouns, verbs, and adjectives are three of the most common POS tags used by developers in crafting identifier names. An instigation on the effectiveness of Stanford Log-linear POS Tagger on field names by Binkley et al. \cite{Binkley2011MSR} shows the need for further refinement of the tagger to support field names.

\subsubsection*{Linguistic Anti-Patterns:} Arnaoudova et al. \cite{Arnaoudova2013CSMR} propose a series of linguistic anti-patterns (i.e., deviation from well-established lexical naming practices), which have been implemented in static analysis-based tools \cite{Arnaoudova2016EMSE,IDEAL}. We envision findings from our study expanding the list of such anti-patterns.

\subsection{Identifier Name Evolution}
Arnaoudova et al. \cite{Arnaoudova2014TSE} present an approach to analyze and classify identifier renamings. The authors also show the impact of proper naming on minimizing software development effort. Peruma et al. utilize this taxonomy in multiple studies as part of their study around contextualizing identifier renaming \cite{Peruma2018IWoR,Peruma2019SCAM,Peruma2020JSSRenames}. The authors show that developer frequently narrow the meaning of the name and observe instances where developers change the plurality of a name in response to its type changing to/from a collection.

The work we present in this paper is primarily complementary to the works discussed above. In particular, understanding the meaning and evolution of digits in identifiers enhances how we interpret the meaning and structure of identifier naming. In turn, this supports stronger analysis of identifier names whether we are using grammar patterns, naming conventions, or other similar techniques to understand the meaning of identifier names. Additionally, this research supports identifier name recommendation, appraisal, and modeling approaches \cite{abebe:2021, tien:2021, IDEAL} due to the fact that a stronger understanding of the meaning of digits can augment the quality of the data used to train and configure these approaches. Finally, poor usage of digits (e.g., overuse) may be considered linguistic anti-patterns. Further study is needed, but the taxonomy we present has the potential to play a significant, supportive role in identifier name quality, appraisal and recommendation research and techniques.

\section{Experiment Design}
\label{Section:experiment_design}
\begin{figure}
 	\centering
 	\includegraphics[trim=0cm 0cm 0cm 0cm, width=1\linewidth]{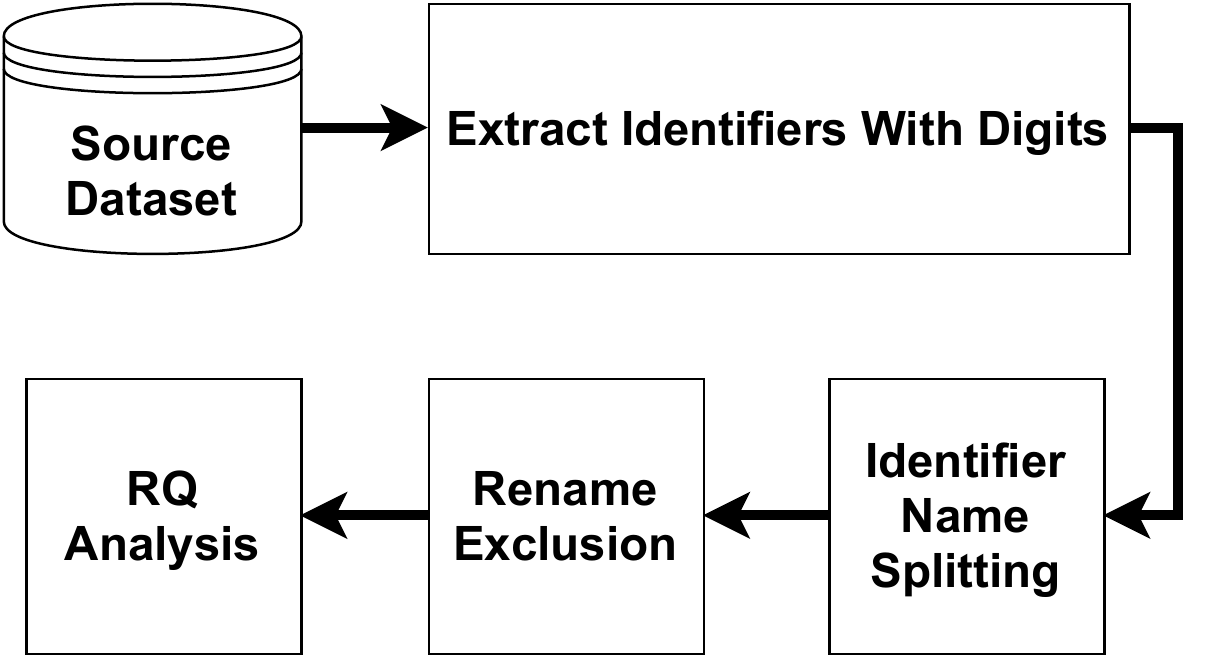}
 	\caption{Overview of our experiment design.}\vspace{-2.5mm}
 	\label{Figure:diagram_experiment}
\end{figure}

This section provides details about the methodology of our study. Depicted in Figure \ref{Figure:diagram_experiment} is an outline of our experiment design. We perform a series of natural language-based processing activities on the source dataset before answering our RQs. In the below subsections, we describe these activities. Furthermore, our dataset is available for replication and extension purposes at \cite{ProjectWebSite}.

\subsection{Source Dataset}
\label{Section:experiment_design_source}
In this study, we utilize the dataset of rename refactorings made available by Peruma et al. \cite{Peruma2020ICPC}. The dataset contains the commit history and refactoring operations of 800 open-source Java systems. The refactoring operations were mined at the commit level using  RefactoringMiner \cite{Tsantalis2018ICSE}, a  state-of-the-art tool with a precision of 98\% and a recall of 87\% \cite{Silva2016FSE, Vassallo2019SCP}, and contain the rename operations performed on classes, attributes, methods, parameters, and local-variables. As shown in Table \ref{Table:refactoring_operations}, in total, the dataset contains 428,079 rename operations and 926,948 non-rename operations. The dataset also indicates if a source code file is a JUnit-based unit test file or not by following an approach similar to \cite{Peruma2019CASCON}.

\begin{table}
\centering
\caption{Volume of rename and non-rename refactoring operations in the source dataset.}\vspace{-2.5mm}
\label{Table:refactoring_operations}
\begin{tabular}{@{}lrr@{}}
\toprule
\multicolumn{1}{c}{\textbf{Refactoring Operation}} & \multicolumn{1}{c}{\textbf{Count}} & \multicolumn{1}{c}{\textbf{Percentage}} \\ \midrule
Rename Attribute                                   & 140,306                            & 10.35\%                                 \\
Rename   Method                                    & 91,235                             & 6.73\%                                  \\
Rename Variable                                    & 89,032                             & 6.57\%                                  \\
Rename   Parameter                                 & 75,042                             & 5.54\%                                  \\
Rename Class                                       & 24,556                             & 1.81\%                                  \\
Move   And Rename Class                            & 7,676                              & 0.57\%                                  \\
Move And Rename Attribute                          & 232                                & 0.02\%                                  \\
\textit{Non-Rename Operations}                     & 926,948                            & 68.41\%                                 \\ \midrule \midrule
\textbf{Total}                                     & 1,355,027                          & 100\%                                 \\ \bottomrule
\end{tabular}\vspace{-2.5mm}
\end{table}

\subsection{Extract Identifiers With Digits}
As our study is limited to analyzing only renames that involve digits in the identifier's name, our initial step is to extract such rename operations from the source dataset. We achieve this by using a regular expression to detect the presence of a digit in either the old or new name of the rename operation; the results of this activity yield 149,980 rename operations.

\subsection{Identifier Name Splitting}
A prerequisite to analyzing identifier names is to split the name into its constituent terms. To perform this activity, we utilize the Ronin splitter algorithm implemented in Spiral, a specialized open-source identifier name splitting Python package  \cite{huckaspiral}. The splitter utilizes heuristic rules, English terms, and token frequencies to determine the individual terms in an identifier's name. Furthermore, prior studies have utilized this package to split identifier names \cite{Peruma2020ICPC,Peruma2020JSSRenames,Peruma2018IWoR}. As an example, the name `service2WsdlResource' is split into`service', `2', `Wsdl', and `Resource'.



\subsection{Rename Exclusion}
\subsubsection{Domain Terms:}
As part of the identifier name splitting activity, the splitter considers standard domain terms that contain digits and refrains from splitting such terms. For example, the splitter splits the name `testi18nGetAll' into the terms `test', `i18n', `Get', and `All' since `i18n' is an established computing abbreviation associated with internationalization and localization. 
Hence, we exclude such rename operations from our analysis. 

\subsubsection{Auto-Generated Code:} A manual examination of the rename instances in our dataset showed the excessive presence of automatically generated code. For example, we encountered identifiers like `LA642\_0' and  `FOLLOW\_EOF\_in\_entryRuleCodeRef1107'. Such code occurs in projects that utilize parser generator frameworks like ANTLR. Identifier names generated automatically are a significant portion of identifiers with numbers in them. Because we wanted to get a good sample of identifiers that included digits but were written by humans, we used heuristics to remove many auto-generated identifiers from our set. To determine how to exclude such source code, we analyzed a statistically significant sample of 383 rename instances (confidence level of 95\% and an interval of 5\%) and determined some common traits we could use to remove them. This helped us get a better, more equal sample that included all identifiers, including some auto-generated identifiers that we were unable to easily exclude. We discuss this more in Section \ref{lastRQ} and as a challenge in Section \ref{Section:discussion}.

\subsection{RQ Analysis}
To answer our RQs, we analyze 15,424 rename operations. This value comprises of 2,683 renames in unit test files and 12,741 renames in non-test files. Furthermore, as shown in Table \ref{Table:reame_identifier}, method variables accounted for most digit-based renames with 28\% instances, followed by method names at 22.98\%. We follow a mixed-methods \cite{Tashakkori1998Mixed} approach to answer our RQs. In this approach, we utilize well-established statistical measures and custom code to report trends and patterns in our dataset and manually analyze the data to present representative examples to complement our quantitative findings. In Section \ref{Section:experiment_results}, where applicable, we elaborate on the methodology of our approach to answer each RQ. We created these research questions with the purpose of understanding both a quantitative and qualitative perspective on the appearance, and purpose, of digits when they appear in identifiers.

\begin{table}
\centering
\caption{Renames distribution by identifier type.}\vspace{-2.5mm}
\label{Table:reame_identifier}
\begin{tabular}{@{}lrr@{}}
\toprule
\multicolumn{1}{c}{\textbf{Identifier Type}} & \multicolumn{1}{c}{\textbf{Count}} & \multicolumn{1}{c}{\textbf{Percentage}} \\ \midrule
Variable                                    & 4,319                            & 28.00\%                                 \\
Method                                     & 3,545                             & 22.98\%                                 \\
Parameter                                       & 2,878                              & 18.66\%                                  \\
Class                                        & 2,779                              & 18.02\%                                  \\
Attribute                                    & 1,903                              & 12.34\%                                  \\\midrule\midrule
\textbf{Total}                                        & 15,424                            & 100\%                                   \\ \bottomrule
\end{tabular}\vspace{-2.5mm}
\end{table}

\section{Experiment Results}
\label{Section:experiment_results}
In this section, we report on the findings of our experiments by answering our RQs. The first RQ is primarily quantitative and examines the structural evolution of the name with respect to the digits in the name. The second RQ examines how digits convey meaning in an identifier's name. Due to space constraints, specific tables in the RQs show only the most frequently occurring instances; the complete set is available in our shared dataset \cite{ProjectWebSite} to encourage replication, future research on this topic, and verification of the results we report in this study.

\subsection*{\RQA}
\vspace{1mm}

\begin{table}
\centering
\caption{Top five distribution of the number of digits in the old and new names.}\vspace{-2.5mm}
\label{Table:digits_in_name}
\begin{tabular}{@{}llrr@{}}
\toprule
\multicolumn{1}{c}{\textbf{\begin{tabular}[c]{@{}c@{}}Number of digits\\ in the old name\end{tabular}}} & \multicolumn{1}{c}{\textbf{\begin{tabular}[c]{@{}c@{}}Number of digits\\ in the new name\end{tabular}}} & \multicolumn{1}{c}{\textbf{Count}} & \multicolumn{1}{c}{\textbf{Percentage}} \\ \midrule

1                                                                          & 1                                                                                   & 5,370                    & 79.93\%                        \\
2                                                                          & 2                                                                                   & 518                    & 7.71\%                        \\
3                                                                          & 3                                                                                   & 207                    & 3.08\%                        \\
1                                                                          & 2                                                                                   & 185                    & 2.75\%                         \\
2                                                                          & 1                                                                                   & 92                     & 1.37\%                         \\
\multicolumn{2}{l}{\textit{Other combinations}}                                                                                                                  & 346                    & 5.15\%                        \\\midrule\midrule
\multicolumn{2}{l}{\textbf{Total}}                                                                                                                               & 6,718                   & 100\%                          \\ \bottomrule
\end{tabular}
\end{table}

\begin{table}
\centering
\caption{Top five distribution of the position of digits in the old and new names.}\vspace{-2.5mm}
\label{Table:position_in_name}
\begin{tabular}{@{}llrr@{}}
\toprule
\multicolumn{1}{c}{\textbf{\begin{tabular}[c]{@{}c@{}}Position of digits\\ in old name\end{tabular}}} & \multicolumn{1}{c}{\textbf{\begin{tabular}[c]{@{}c@{}}Position of digits\\ in new name\end{tabular}}} & \multicolumn{1}{c}{\textbf{Count}} & \multicolumn{1}{c}{\textbf{Percentage}} \\ \midrule
2\textsuperscript{nd}                                                                                                   & 2\textsuperscript{nd}                                                                                                   & 1,930                             & 28.73\%                                 \\
3\textsuperscript{rd}                                                                                            & 3\textsuperscript{rd}                                                                                            & 757                              & 11.27\%                                  \\
4\textsuperscript{th}                                                                                                  & 4\textsuperscript{th}                                                                                                  & 432                              & 6.43\%                                  \\
5\textsuperscript{th}                                                                                                   & 5\textsuperscript{th}                                                                                                   & 328                              & 4.88\%                                  \\
2\textsuperscript{nd}                                                                                                   & 3\textsuperscript{rd}                                                                                                   & 283                              & 4.21\%                                  \\
\multicolumn{2}{l}{\textit{Other combinations}}                                                                                                                                                               & 2,988                            & 44.48\%                                 \\ \midrule\midrule
\multicolumn{2}{l}{\textbf{Total}}                                                                                                                                                                            & 6,718                            & 100\%                                   \\ \bottomrule
\end{tabular}
\end{table}

\begin{table}
\centering
\caption{The most frequently occurring instance of digit position and count in test and non-test files.}\vspace{-2.5mm}
\label{Table:test_notest}
\resizebox{\columnwidth}{!}{%
\begin{tabular}{@{}clrr@{}}
\toprule
\textbf{\begin{tabular}[c]{@{}c@{}}Number of digits\\ in the old name\end{tabular}} & \multicolumn{1}{c}{\textbf{\begin{tabular}[c]{@{}c@{}}Number of digits\\ in the new name\end{tabular}}} & \multicolumn{1}{c}{\textbf{Count}} & \multicolumn{1}{c}{\textbf{Percentage}} \\ \midrule
\multicolumn{4}{c}{\textit{Test Files}}                                                                                                                                                                                                                                      \\
\multicolumn{1}{l}{1}                                                               & 1                                                                                                       & 1,126                              & 82.73\%                                 \\
\multicolumn{4}{c}{\textit{Non-Test Files}}                                                                                                                                                                                                                                  \\
\multicolumn{1}{l}{1}                                                               & 1                                                                                                       & 4,244                              & 79.22\%                                 \\ \midrule
\textbf{\begin{tabular}[c]{@{}c@{}}Position of digits\\ in old name\end{tabular}}   & \multicolumn{1}{c}{\textbf{\begin{tabular}[c]{@{}c@{}}Position of digits\\ in new name\end{tabular}}}   & \multicolumn{1}{c}{\textbf{Count}} & \multicolumn{1}{c}{\textbf{Percentage}} \\ \midrule
\multicolumn{4}{c}{\textit{Test Files}}                                                                                                                                                                                                                                      \\
\multicolumn{1}{l}{2\textsuperscript{nd}}                                                               & 2\textsuperscript{nd}                                                                                                       & 283                                & 20.79\%                                 \\
\multicolumn{4}{c}{\textit{Non-Test Files}}                                                                                                                                                                                                                                  \\
\multicolumn{1}{l}{2\textsuperscript{nd}}                                                               & 2\textsuperscript{nd}                                                                                                       & 1647                               & 30.74\%                                 \\ \bottomrule
\end{tabular}%
}

\end{table}

\begin{table*}
\centering
\caption{Top five distribution of the position of digits and the number of digits in the old and new names.}\vspace{-2.5mm}
\label{Table:digits_positions}
\resizebox{\textwidth}{!}{%
\begin{tabular}{@{}llllrr@{}}
\toprule
\multicolumn{1}{c}{\textbf{Position of digits in old name}} & \multicolumn{1}{c}{\textbf{Position of digits in new name}} & \multicolumn{1}{c}{\textbf{Number of digits in the old name}} & \multicolumn{1}{c}{\textbf{Number of digits in the new name}} & \multicolumn{1}{c}{\textbf{Count}} & \multicolumn{1}{c}{\textbf{Percentage}} \\ \midrule
2\textsuperscript{nd}                                                        & 2\textsuperscript{nd}                                                         & 1                                                             & 1                                                             & 1,930                              & 28.73\%                                 \\
3\textsuperscript{rd}                                                        & 3\textsuperscript{rd}                                                         & 1                                                             & 1                                                             & 757                                & 11.27\%                                 \\
4\textsuperscript{th}                                                        & 4\textsuperscript{th}                                                         & 1                                                             & 1                                                             & 432                                & 6.43\%                                  \\
5\textsuperscript{th}                                                        & 5\textsuperscript{th}                                                         & 1                                                             & 1                                                             & 328                                & 4.88\%                                  \\
2\textsuperscript{nd}                                                        & 3\textsuperscript{rd}                                                         & 1                                                             & 1                                                             & 283                                & 4.21\%                                  \\
\multicolumn{4}{l}{\textit{Other Combinations}}                                                                                                                                                                                                          & 2,988                              & 44.48\%                                 \\ \midrule\midrule
\multicolumn{4}{l}{Total}                                                                                                                                                                                                                                & {6,718}          & {100\%}               \\ \bottomrule
\end{tabular}%
}\vspace{-2.5mm}
\end{table*}

In this RQ, we take a quantitative approach to study the treatment of names with digits over time. To this extent, we examine the presence and/or absence of digits in an identifier's name before and after a rename operation. Findings from this RQ will give insight into the prevalence of digits, the digits frequently utilized in an identifier's name, and the action taken on these identifiers.

From the set of 15,424 renames with digits, we observe that 6,718 (or 43.56\%) instances have a digit present in the old and new name (e.g., `h1' $\rightarrow$ `gain1'). Renames where all digits are removed from the name account for 5,135 (or 33.29\%) instances (e.g., `arg1' $\rightarrow$ `id'), while  3, 571 (or 23.15\%) instances show the adding of digits to a name that did not contain digits (e.g., `log' $\rightarrow$ `log1'). 

Our analysis now focuses on the instances where digits are preserved in a rename. First, when comparing the number of digits in the old and new name, we observe that 6,170 or 91.35\% instances have an equal number of digits in the old and new name. Additionally, from this list, we observe that most instances (5,370, or 79.93\%) contain only one digit in the old and new names. The next highest set of names contains two digits in the old and new name with 518 or 7.71\% instances (e.g.,  `april7th2011' $\rightarrow$ `april8th2011'). Table \ref{Table:digits_in_name} shows the distribution of the top five combinations of the number of digits in old and new names. Our subsequent analysis looks at the position of the digits occurring in the names. Findings show that most renames (i.e., 4,402 or 65.53\% instances) preserve the position of the digits in the new name. Furthermore, a digit will likely appear in the second position of the old and new names (e.g., `node2' $\rightarrow$ `node3'). 

Table \ref{Table:position_in_name} shows the distribution of the top five combinations of the positioning of digits in the old and new names. Furthermore, when comparing renames in test and non-test files, the most occurring rename combination regarding the number of digits and digit position is the same for both file types. Table \ref{Table:test_notest}, shows the frequently occurring instance for each file type. Finally, a combination of digit count and position (refer Table \ref{Table:digits_positions}) shows that identifier names typically use a single digit that appears as the second term in the old and new name; we observe 1,930 or 28.73\% instances of this pattern combination. We also observe that the single-digit occurrence in the names frequently happens regardless of the position of the digit. 

Our final analysis, in this RQ, looks at the value of the digit appearing in the name. Looking at renames with digits in both names, our observations show that in most rename instances, the value `2' is present in the old and new names and is the only digit present in both names (e.g., `slave2Index' $\rightarrow$ `channel2Index'). This specific occurrence accounts for 1,052 or 15.56\% instances. The next highest occurrence is the value `1' in 851 rename instances as the only digit in the old and new name. Next, examining rename instances where at least one name contains a digit, we observe that the value `1' is either removed from the old name (e.g., `arg1' $\rightarrow$ `id') or added to the new name (e.g., `oldDelta' $\rightarrow$ `delta1') frequently with 1,563 and 1,512 instances, respectively. Finally, our observations show a diverse set of numeric values utilized in naming identifiers, resulting in a long tail when examining the frequency occurrence. 

\textbf{Summary for RQ1.} When renaming identifiers containing digits, a frequent activity is preserving digits in the new name and usually using the same number of digits in the new name and preserving the position of the digits in the name. Furthermore, a single digit in the old and new name is the most commonly occurring pattern, with the digit usually occurring as the second term in the old and new name; a phenomenon common to test and non-test identifier names. Additionally, identifiers utilize a diverse set of digits in their name, with the number `2' being a common digit in the old and new name.

\subsection*{\RQC}\label{lastRQ}
\vspace{1mm}
The prior RQ was primarily quantitative-based, showing how digits evolve in an identifier's name. However, while the findings can contribute to the appraisal of identifier names, we do not know why developers utilize digits in identifiers and how these digits contribute to the overall meaning of the identifier. Hence, in this RQ, we manually examine the semantics of identifiers that include digits, and their surrounding source code, to determine the rationale for the presence of digits. This RQ aims to produce a taxonomy showing how digits convey meaning in an identifier's name. Our analysis is constrained to 375 rename instances, where either the old, new, or both names contain a digit. Furthermore, this dataset represents a stratified statistically significant sample; we utilized a confidence level of 95\% and an interval of 5\% for each identifier type (i.e., class, attribute, method, local-variable, and parameter). In the annotation process, the two authors annotated the dataset with the rationale for the presence of the digit in the name. This was done by reading the identifier name and looking at the code associated with the identifier. Once this was complete, the authors compared their annotations and resolved any conflicts through discussion. Additionally, in specific instances where we encounter interesting phenomena, we perform a snowballing activity to locate examples of additional instances in the original dataset. We discuss taxonomy below and provide specific definitions in Table \ref{Table:taxonomy}.

\begin{table*}
\centering
\caption{Taxonomy showing how digits convey meaning in an identifiers name.}\vspace{-2.5mm}
\label{Table:taxonomy}
\resizebox{\textwidth}{!}{%
\begin{tabular}{@{}lll@{}}
\toprule
\multicolumn{1}{c}{\textbf{Category}} & \multicolumn{1}{c}{\textbf{Definition}} & \multicolumn{1}{c}{\textbf{Example}} \\ \midrule
\textbf{Auto-Generated} & \begin{tabular}[c]{@{}l@{}}Identifier names in this category are generated by a code generation tool, framework, or IDE. The \\ number in these identifiers may have significant meaning to the technique that was used to generate \\ them. However, it is difficult to understand this meaning (if it exists) without a thorough understanding \\ of the technique used to generate them. These identifiers are also typically not written in a way to \\ support comprehension, since they are not typically maintained by developers directly but, instead, \\ regenerated by the technique that created them in the first place.\end{tabular} & \textit{LA18\_6} \\ \midrule
\textbf{Distinguisher} & \begin{tabular}[c]{@{}l@{}}Identifier names in this category differ only by a digit, which is typically appended as the rightmost token.\\ There should be at least two identifiers having a lexically identical name. The purpose of the digit is to avoid\\ name collision at compile/parse time. The digit has no other significant meaning relating to the purpose of\\ the identifier. Thus, the digit primarily operates to distinguish the identifier it is part of from other\\ identifiers that are lexically identical other than their own digit.\end{tabular} & \textit{auditLog3} \\ \midrule
\textbf{Synonym} & \begin{tabular}[c]{@{}l@{}}Identifier names in this category contain at least one digit utilized in place of a word. Sometimes digits \\ are used to represent the meaning typically associated with certain words. The numbers 2 and 4 are very \\ common examples of this, with 2 being associated with words like 'to' and 'too' while 4 is associated with \\ words like 'for'. In this way, they function as a type of shorthand.\end{tabular} & \textit{convert2RList} \\ \midrule
\textbf{Version Number} & \begin{tabular}[c]{@{}l@{}}Identifier names in this category contain at least one digit used to signify a version number. This typically\\ means that the identifier represents an entity whose version is significant to its capabilities and limitations \\ Version numbers were often used in the dataset to inform developers of what version of a framework, \\ tool, protocol, etc, an identifier represented. This could, for example, be an identifier that represents an \\ HTTP 1.0 request having the number 1\_0 appended to it.\end{tabular} & \textit{V1DozerTransformModel} \\ \midrule
\textbf{Specification} & \begin{tabular}[c]{@{}l@{}}Identifier names in this category contain at least one digit that represents a known specification. In many\\ cases, this is a number that acts as a way to uniquely identify concepts, behaviors, or characteristics. These \\ can be mathematical concepts, such as using 3D in identifiers that deal with 3-dimensional data;\\ documented, project-specific behavior like in filter1\_2, where the 1\_2 in the identifier tells us which \\ specific filter (i.e., 1\_2) this identifier represents; and data characteristics such as 9 in arialRegular9Dark\\ which gives us the size of the font data associated with the identifier.\end{tabular} & \textit{arialRegular9Dark} \\ \midrule
\textbf{Domain/Technology} & \begin{tabular}[c]{@{}l@{}}Identifier names in this category have a digit that is part of the name of a domain term or technology. The\\ digits themselves have no individual meaning besides the meaning endowed by the technology/domain that \\ they are a reference to.\end{tabular} & \textit{resultDoubleExp4j} \\ \bottomrule
\end{tabular}
}\vspace{-2.5mm}
\end{table*}

\subsubsection*{\textbf{Auto-Generated}:}
Identifiers falling under this classification are of two types - 1) part of a source file that is entirely auto-generated by a tool/framework (e.g., ANTLR) and 2) specific identifiers generated by the IDE in a source file containing developer-defined identifiers. The identifier names in tool/framework generated source files sometimes have the name of the data type or expression statement. However, like the integer variable `LA18\_6', this is not always the case. In the case of the latter, these identifiers are typically user interface (UI) controls. When a developer utilizes the IDE to drag-and-drop a UI control, such as a textbox,  the IDE generates the code associated with the UI control (i.e., properties and event handler). The name of such identifiers starts with the type of control and ends with an incrementing digit. For example, `jComboBox2' is the IDE generated name for a JComboBox UI control; the value `2' indicates that this is the second JComboBox control in the class. We also observe instances where developers rename auto-generated identifier names to more meaningful names, like \seqsplit{`jScrollPane4' $\rightarrow$ `scrollCompilerDescription'}. The fact that some of these auto-generated names eventually received a higher-quality name indicates that auto-generated identifiers should not necessarily be discarded when performing this kind of research. In some cases, despite being auto-generated, they provide additional insight. The presence of these identifiers represents an opportunity for future research to study how developers use and modify auto-generated identifiers and how they differ, in terms of how they evolve, from other types of identifiers that originated from humans instead of code-generation technologies.

\subsubsection*{\textbf{Distinguisher:}}
These types of identifier names are prevalent in our dataset. Developers create multiple identifiers in the source file with the same name, but each has a unique digit to avoid name collision at compile-time. Hence, the purpose of digits in these names is to distinguish them, lexically, from one another-- hence their name. These types of names were noted but not studied in other research \cite{Newman2019ICSMEa}. In most cases, the digit is the last term in the name, and developers increment the value of the digit with each new lexically-identical iteration of the identifier. An example of such an instance is the declaration of the variables `auditLog1', `auditLog2', and `auditLog3', within the same method, where the digit is a distinguisher. A variant of this category is the use of generic names (and sometimes abbreviations) for identifiers. For example, we encounter method parameters named `arg0' and `arg1'.

\subsubsection*{\textbf{Synonym}}
Developers utilize digits as synonyms for prepositions, with the value `2' frequently used as a substitute for the `to' preposition. One such use of this digit is to indicate the result of an action or a process, such as naming transformation-based identifiers. Such identifiers either convert data from one format to another or hold the transformation results. For example, the method `convert2RList' converts a table-based object, passed as a parameter, to a list-based object which it returns. Another use of the digit `2' is to indicate purpose or intention. For example, an attribute with the name `mb2use' is utilized to hold the cache size that the program intends to use. We also encounter the use of the value `4' as a synonym for the preposition `for' as in the method `populate4Test', which instantiates a variable for use in test cases. Furthermore, such digits are contained within the name and not at the end of the name.  

\subsubsection*{\textbf{Version Number}}
Developers utilize identifiers to signify the version number the code supports. For example, looking at the comments in the class `V1DozerTransformModel', the term `V1' represents ``A version 1 DozerTransformModel''. A drawback of such an approach is the continual update of the identifier's name with new versions of the system, as shown by the renaming of the variable `pg75'  $\rightarrow$ `pg80'. However, certain developers also acknowledge this overhead and remove such dependencies as in the renaming of the identifier `DROID\_SIGNATURE\_FILE\_V45' $\rightarrow$ `DROID\_SIGNATURE\_FILE' with the commit message ``Changed variable name so that it is more generic than previously''

\subsubsection*{\textbf{Specification}}
The digits in a name can relate to specifications of the system,  such as size or dimensions, or provide details about the system's behavior. For instance, the value `9' in the attribute name `arialRegular9Dark' indicates the size of the font object associated with the attribute. In another example,  the comments for the class \seqsplit{`Geographic3DTo2DConversion'} show that this class ``Converts between a Geographic 3D and a 2D system''. Though rare, we also encounter the use of digits to specify a specific use case, as in the name \seqsplit{`testUC\_3\_EraseFacetClassifier\_NoSource'}; this is a unique traceability mechanism, especially for test cases

\subsubsection*{\textbf{Domain/Technology}}
The presence of digits is also due to developers using domain or technology names that contain digits. For instance, we observe developers utilizing  API/library names as a term in the identifier's name, like `Twitter4J', `Neo4J', `Slf4j', `Log4j' , `Args4j',  Exp4j, and `Junit4', such as in naming the attribute`resultDoubleExp4j'. The use of standards/formats in the name also results in digits in the name. For example, in the method name `testCP437FileRoundtrip', the term `CP437' is a standard for character encoding. Likewise, the value `1516' in \seqsplit{`parseInvalidHla1516eFomWithUndefinedTransportForAttribute'}  corresponds to an IEEE standard (i.e., IEEE-1516e).

\textbf{Summary for RQ2.}
There are multiple reasons why an identifier name contains digits. Digits are frequently utilized to create unique identifier names; this is true for both auto-generated and developer-written source code. Developers also name identifiers based on domain and technology terms such as the API/library associated with the identifier, resulting in digits if the API/library name contains such values. Additionally, digits are also utilized to convey size, standards, versions, and words like 'to' and 'for'.

\section{Threats To Validity}
\label{Section:threats}
Our sample consists of open-source Java systems that were selected from a dataset of engineered projects, meaning that the projects we studied all followed well-known software development practices. Thus, while the findings in this paper are limited to Java systems, there is some likelihood that some of what we discuss extends to projects written in other languages-- further study is needed to determine how well our results generalize.

Our initial extraction of digit-based identifiers had a significant amount of auto-generated code. We saw this as a threat because part of the goal of this study is to understand how humans use digits in code; not just code generation techniques. Thus, we re-extracted after using heuristics to remove a large amount of the auto-generated code. This did not remove all auto-generated code. While having all types of identifiers that include digits represented in our set is advantageous for this study's goal, it was not possible for us to completely balance our dataset because it was not possible, in general, to determine if an identifier is auto-generated or not before manually examining it. Thus, while the re-extracted dataset is more representative of all identifiers, it may still be biased in favorite of human-create or machine-created identifiers. In addition, our RQ2 annotation sample was statistically significant with a 95\% confidence level and a 5\% confidence interval. Thus, even if it is biased, it is representative (within those boundaries) of the types of identifiers present in the population. This sample was also stratified in terms of identifier type (i.e., attribute, method, class, local-variable, and parameter).

The construction of the taxonomy presented in RQ2 was done using the sample described above. The authors individually annotated identifiers within the dataset by annotating the purpose of the digit found within the identifier. They then reviewed one anothers' annotations and agreed on the purpose of each digit they reviewed based on one another's feedback and on the code context surrounding the identifier. While this process does not guarantee that the categories within our taxonomy are complete, it helps ensure that the categories defined in the taxonomy are valid. Further research can mitigate the threat that this taxonomy or the definitions found within are incomplete using the dataset we make openly available.

Finally, our methodology utilizes specific tools to process the identifiers. Each of these tools introduces a threat in that they may not be completely accurate (e.g., the splitter could make a mistake). However, these tools are well-established and used in similar work. They represent the state-of-the-art in their individual domains, or have been shown to be extremely accurate.

\section{Discussion \& Conclusion}
\label{Section:discussion}
In this exploratory paper, we examine the presence of digits in identifier names and their evolution by studying the renaming of identifiers. Our findings show that developers consider digits as essential elements in an identifier's name and utilize such digits to convey a specific purpose. Further research is required to understand whether the uses of digits that we have outlined in our taxonomy are good or bad practices. Additionally, our findings add to the body of knowledge of identifier naming practices and can be utilized in future implementations of identifier name appraisal and recommendation tools. Specifically, our results provide hints for these tools to use to determine when the use of a digit helps or hurts comprehension and for determining why a digit is present within a given identifier. Below, we discuss how the findings from our RQs support the community through a series of takeaways.


\textbf{Takeaway 1 - \textit{Digits Are Preserved Post-Rename.}} The renaming of identifiers containing digits typically preserves the digits. These findings better equip the development of the tools and techniques to improve identifier name recommendations and appraisals when developers perform rename operations. 
The utilization of such characteristics in name appraisal tools increases the appraisal accuracy and provides the developer with an explanation of the appraisal results. Furthermore, our observations show the association of digits with generic terms, as in the case of `parameter0' or the abbreviation `arg1'. Such occurrences are examples of poor-quality names since the names do not indicate the purpose of the identifier. It is interesting to note that such names are usually developer-generated and not auto-generated. An examination of the commit history shows developers introducing such names early in the file's lifetime. The use of appraisal tools during implementation can prevent the occurrences of such phenomena. In another similar case, the name crafted by the developer contains the name of the datatype (e.g., \texttt{int int1} and \texttt{int[] arrayOfInt1}), which is generally considered poor naming practice \cite{martin2009clean}.

\textbf{Takeaway 2 - \textit{The Digits Found in Identifier Names Are Meaningful.}} RQ2 presents a taxonomy of how digits in an identifier convey meaning. While most identifier names are either autogenerated or utilized as distinguishers, we see specific instances where there is a relationship between the digits and the code/behavior it represents, such as when used as synonyms for words. These findings are helpful in multiple scenarios. For instance, name appraisal tools can utilize static analysis techniques to determine if the digit present in the name is related to the code, as in the case of linguistic anti-patterns \cite{Arnaoudova2014TSE}. Furthermore, the digits serve to build a catalog of technologies, standards, or domain terms utilized in the project, which serve as input into project and code summarization. Finally, while correcting the identifier names in source files that are entirely auto-generated might seem redundant since they may be regenerated, name appraisals and recommendation tools can focus on distinguisher-based names in source files containing a mix of system and developer generated code. 

\textbf{\textit{Challenges}}. While auto-generated identifiers frequently contain digits, we also wanted to see how human developers use digits in their identifier naming practices. Determining when an identifier is auto-generated is challenging to do automatically, and while we discovered some heuristics that we could use to reduce the number we collected, we did not create a general method for detecting them. While removing auto-generated identifiers is not always appropriate, detecting them in a generalized manner would support future work by making it significantly easier to separate them. However, as noted in Section \ref{lastRQ}, we did find cases where developers generated code/identifiers automatically and then went back and renamed those identifiers to be more readable. These are interesting cases and present another opportunity-- if we can find a way to detect auto-generated identifiers automatically, it becomes easier to study them when developers later rename them. This could provide some novel insight into naming practices, since we would be able to analyze these identifiers to understand more about how developers interact with them in contrast to developer-made identifiers.

\section{Future Work \& Acknowledgements}
Digits within an identifier's name exist for a specific purpose and can be used to understand more about the behavior and information being conveyed by an identifier to developers. We believe this data can assist in providing more robust identifier name recommendations and appraisals by improving our understanding of how digits are, and should be, used to convey information.

This material is based upon work supported by the National Science Foundation under Grant No. 1850412.

\bibliographystyle{ACM-Reference-Format}
\bibliography{references}

\end{document}